\documentclass[12pt]{article}

\usepackage{scicite}
\usepackage{graphicx}
\usepackage{times}
\usepackage[version=3]{mhchem} 
\usepackage{color}

\newenvironment{sciabstract}{%
\begin{quote} \bf}
{\end{quote}}

\title{Probing the Ultimate Plasmon Confinement Limits with a Van der Waals heterostructure}

\author{David Alcaraz Iranzo,$^{1\dagger}$ S\'{e}bastien Nanot,$^{1,2\dagger}$ Eduardo J. C. Dias,$^{3}$ \\
Itai Epstein,$^{1}$ Cheng Peng,$^{4}$ Dmitri K. Efetov,$^{1,4}$ Mark B. Lundeberg,$^{1}$\\
Romain Parret,$^{1}$ Johann Osmond,$^{1}$ Jin-Yong Hong,$^{4}$ Jing Kong,$^{4}$ \\
Dirk R. Englund,$^{4}$ Nuno M. R. Peres,$^{3}$ Frank H.L. Koppens,$^{1,5\ast}$
\\
\normalsize{$^{1}$ICFO - The Institute of Photonic Sciences, The Barcelona Institute of Science and }\\
\normalsize{Technology, 08860 Castelldefels (Barcelona), Spain}\\
\normalsize{$^{2}$L2C - Laboratoire Charles Coulomb, Universit\'{e} de Montpellier, CNRS, }\\
\normalsize{34095 Montpellier Cedex, France}\\
\normalsize{$^{3}$Centro de F\'{i}sica and Departamento de F\'{i}sica and QuantaLab, Universidade do Minho, }\\
\normalsize{P-4710-057 Braga, Portugal}\\
\normalsize{$^{4}$Department of Electrical Engineering and Computer Sciences, Massachusetts }\\
\normalsize{Institute of Technology, Cambridge, Massachusetts 02139, United States}\\
\normalsize{$^{5}$ICREA – Instituci\'{o} Catalana de Recerca i Estudis Avan\c{c}ats, Barcelona, Spain}\\
\normalsize{$^\ast$To whom correspondence should be addressed; E-mail:  frank.koppens@icfo.eu.}\\
\normalsize{$^{\dagger}$These authors contributed equally to this work.}
}

\date{}

\begin{document} 


\baselineskip24pt


\maketitle


\begin{sciabstract}
The ability to confine light into tiny spatial dimensions is important for applications such as microscopy, sensing and nanoscale lasers. 
While plasmons offer an appealing avenue to confine light, Landau damping in metals imposes a trade-off between optical field confinement and losses. We show that a graphene-insulator-metal heterostructure can overcome that trade-off, and demonstrate plasmon confinement down to the ultimate limit of the lengthscale of one atom. This is achieved by far-field excitation of  plasmon modes squeezed into an atomically thin hexagonal boron nitride dielectric h-BN spacer between  graphene and metal rods. A theoretical model which takes into account the non-local optical response of both graphene and metal is used to describe the results. These ultra-confined plasmonic modes, addressed with far-field light excitation, enables a route to new regimes of ultra-strong light-matter interactions.\end{sciabstract}

%


Van der Waals heterostructrures are constructed by vertically stacking atomically thin materials, selected from a rich palette of thousands of materials such as graphene (semi-metal), hexagonal boron nitride (h-BN, dielectric) and transition metal dichalcogenides (semi-conductors) \cite{geim2013}. These are key enablers for tailoring electronic, optical and opto-electronic properties \cite{Novoselov2016}.
The most common heterostructure for 2D electronics is graphene encapsulated by h-BN, and recently this system has also emerged as a platform for polaritonics \cite{low2017polaritons,basov2016polaritons}, with the capability to strongly confine plasmon polaritons with a relatively long plasmon lifetime exceeding 500 fs at room temperature\cite{Woessner2014}. Heterostructures of graphene, h-BN, and metals have revealed so-called propagating acoustic plasmons\cite{principi2011,Popov2011,Alonso-Gonzalez2017}, in which metal screening confines the light in the space between the metal and the graphene,  and it slows down the plasmon to a velocity almost as low as $c$/300 (with $c$ the speed of light) \cite{Lundeberg2017}. What is then the ultimate limit on the  confinement of propagating or resonant plasmons?  For bulk metal-based plasmonic systems, such as tapers \cite{Stockman2004}, grooves \cite{Bozhevolnyi2005}, metal-insulator-semiconductor \cite{Oulton2008}, metal-insulator-metal\cite{Dionne2006,miyazaki2006} waveguides, the confinement of propagating surface plasmon polaritons is limited by Landau damping \cite{bozhevolnyi2016quantum}. For example, the quality factor of plasmonic Fabry-P\'erot modes dropped below one for a confinement below 15 nm\cite{miyazaki2006}. Further enhancement of optical fields to the nanometer scale is possible in hotspots \cite{Ciraci2012,Benz2016,teperik2013robust}, although these are broadband in character.

We present a  graphene-insulator-metal platform that allows us to realize and probe the ultimate physical limits of (out-of-plane) confinement of propagating plasmons down to the ultimate physical boundary of one atom-thick layer (here $\lambda_0/26000$), and without sacrificing damping. We take advantage of the fact that plasmons in two-dimensional materials are fundamentally different from plasmons in bulk metals as the restoring force by the long-range Coulomb interactions, essential for the plasmon confinement,  can be controlled by tailoring the external environment. For that reason the out-of-plane confinement and wavelength compression can be increased strongly without suffering from Landau damping. We use far-field light to couple to these strongly confined plasmons and find a vertical mode length down to 0.3 nm, while higher order Fabry-P\'erot resonances reveal that the propagating character of the plasmons is preserved.

The basic device geometry consists of graphene as the plasmonic material, encapsulated by atomically thin dielectric materials (h-BN or \ce{Al2O3}) and covered by a metallic rod array (Fig. 1A) (see \cite{sup2} for details on the fabrication processes). The advantages of the periodic metal-insulator-graphene system is twofold.
 First, the presence of the metal results in efficient screening of the graphene plasmons, squeezing the so-called screened graphene plasmons (SGPs) into the graphene-metal gap without reducing their lifetime. Although screening and coupling of radiation to plasmons in classical 2D electron gases (2DEG) was previously achieved using grating-gate field-effect-transistors in the THz range \cite{Popov2011}, a relatively thick barrier layer ($\sim100~$nm) prevented the confinement of plasmons to the sub-nanometer limit, as we report here. Second, metal rods facilitate efficient coupling between far-field light and the strongly confined plasmons, where the width of the rods define the resonant conditions for the plasmon modes. This approach does not require patterning of the graphene  into nanoribbons or nanodiscs as for previous infrared graphene plasmonic studies with far-field light \cite{Brar2013,Yan2013c}. 

The effect of the coupling of far-field light into our devices is represented by the field profiles obtained by Finite-Difference-Time-Domain (FDTD) simulations  (Figs. 1B, C).  The plasmons are launched at the metal edges and most of the electric field is confined between metal and graphene, with virtually no leakage into the metal. This confinement arises because the metal acts as a nearly perfect conductor that prevents (most of the) field penetration. As image charges are induced in the metal, the plasmon mode is analogous to the anti-symmetric plasmon mode of two nearby graphene sheets with twice the spacer thickness\cite{Alonso-Gonzalez2017}, known as acoustic plasmons, carrying larger momentum than for conventional plasmon resonances in graphene ribbons (see  \cite{sup2}, section 5.7).
Within the dielectric gap, the plasmons maintain their propagating character and reflect at the edges of the rod, forming what appears to be a standing wave pattern similar to two coupled  Fabry-P\'erot resonators\cite{Lin2017}. 

Bringing the metal closer will increase the plasmon screening, which slows down the plasmons as previously observed with s-SNOM by a proximity to a metal layer\cite{Alonso-Gonzalez2017,Lundeberg2017}, and also enhances the vertical confinement, as shown for two different materials and graphene conductivity models (local and non-local) (Fig. 1D).  The most extreme case is that of  a monolayer h-BN spacer, where in theory the vertical plasmon confinement is below 1~nm.  Interestingly,  the calculated width of the resonance (shown in \cite{sup2}, section 6) does not increase when reducing the spacer thickness $s$.
Therefore, this platform allows us to access the ultimate confinement limits of propagating plasmons in two spatial directions: out-of-plane confinement defined by $s$ and in-plane confinement governed by $\lambda_0/\lambda_p$.  

%
%
%

The far-field approach presented above allows probing of this plasmon confinement using Fourier Transform Infrared (FTIR) transmission measurements. Gate dependent, spectral extinction ($1-T/T_{\rm{CNP}}$) curves are measured (Fig. 2A) for a device with continuous graphene (unpatterned), covered with  $2$~nm \ce{Al2O3} spacer ($s=2$ nm) and metallic rods of 256 nm ($w=256$ nm). These curves are obtained from the transmission curves $T$ normalized by the transmission of undoped graphene $T_{\rm{CNP}}$ (at the charge neutrality point, CNP) for several gate voltages (hence, Fermi energies $E_{\rm{F}}$, see \cite{sup2} section 2.2) at the same position, with light polarized perpendicular to the rods long axis.
 The spectra in Fig.~\ref{fig:meas1}a exhibit multiple resonances, which become more pronounced and continuously blue shift with increasing Fermi energies, thus confirming their plasmonic nature. 
The appearance of multiple peaks demonstrates that the incoming light can resonantly couple to higher orders of plasmonic modes than reported previously using nanoribbons\cite{Yan2013c}. We find up to five visible resonances that can be controlled by the spacer thickness and the metallic array geometry.

We corroborate the phenomenon by examining the simulated electric field intensity profiles at the graphene surface for the same geometry as the measured device (Fig.~\ref{fig:meas1}B) and the corresponding simulated extinction spectra (Fig.~\ref{fig:meas1}C).
The simulations, obtained by FTDT simulations (see \cite{sup2} section 4), show a good agreement with the experiment (for $s$=2nm) in terms of shape and peak position using the local optical response of graphene as well as a local metal permittivity models. 
The resonances can be related to Fabry-P\'erot behavior of the propagating SGPs below the metal and the resonance orders $m$ can be approximated by the half number of nodes in the field profile (Fig.~\ref{fig:meas1}B).
This observation confirms that resonances up to the $7^{th}$ order are contributing to optical extinction, while for patterned graphene \cite{Yan2013c,Fang2013} only very weak second order resonances have been experimentally reported.
We attribute the efficient launching of these higher-order resonances to the strong dipole modes at the metal edges (\cite{sup2} section 9), which provide the required momentum to scatter light into graphene plasmons.

To probe the physical limits  of SGP confinement we studied devices with a metal-graphene spacer with only one monolayer of CVD-grown h-BN of thickness 0.7nm \cite{Kim2012,Park2014}. 
The extinction spectra were obtained for various $E_{\rm F}$ of a device with $w=33$~nm  and $g=37$~nm (Fig. 3A).
Strikingly, even in this case, the SGP resonances are still visible with high extinction values. 
A single plasmon peak  is clearly visible and blue shifts for increasing  $E_{\rm F}$, while hybridizing and anti-crossing with \ce{SiO2} and \ce{h-BN} phonons. This shift with Fermi energy confirms that the extinction resonance is due to graphene plasmons. 

The propagating character of the SGP modes can be assessed by further increasing  $w$, which allows us to probe higher order Fabry-P\'erot resonances.
We will show that this measurement is equivalent to changing the plasmonic cavity width $w$ which defines the resonant conditions.
Extinction of several devices with increasing $w$ and similar gap around $g=40$~nm and doping of $E_{\rm F}=0.54$~eV were studied (Fig. 3B).
Additional doping dependences of these devices can be found in Fig.S5.
The first order resonance (indicated by the red dashed lines in Fig~\ref{fig:meas2}B)  displays hybridization with the optical phonons of its dielectric environment (as observed for plasmons in graphene nanoribbons \cite{Brar2014}).
We observe that when increasing $w$, the first order resonance shifts to lower energies and hybridizes not only with the h-BN phonons, but also with the \ce{SiO2} phonons. 
More resonances appear with increasing $w$.
For example, for $w=64~nm$, a second order resonance appears at 1900~cm$^{-1}$ (green dashed line), which also shifts for larger $w$ and starts hybridizing when increasing $w$.
For the largest $w$, the $3^{rd}$ (pink dashed line) and even $4^{th}$ order resonances appear for our measurement range.

The same trends are seen in the calculated spectra (Fig. 3C), obtained from a semi-analytical approach, which consists of a Fourier decomposition of the fields (for TM-modes) in each dielectric region of the system (see \cite{sup2} section 5).
In addition, for these very tightly confined optical fields one must also take into account that the non-local optical response of both the metal and graphene can have appreciable effects. First of all, the in-plane momentum of the graphene plasmon is strongly enhanced by the presence of the metal and approaches $\omega/v_{\rm{F,graphene}}$, where the momentum dependence of the graphene optical conductivity (i.e. non-local corrections to the conductivity) increases \cite{Lundeberg2017}.    These graphene non-local corrections are modelled within the framework of the random-phase-approximation (RPA, see \cite{sup2} section 5.6).
Second, due to additional strong vertical field confinement of the order of 1~nm, the components of out-of-plane wavevectors approach $\omega/v_{F,metal}$, which is the Fermi momentum of the charge carriers in the  metal\cite{Ciraci2012}. This trend can lead to additional non-local effects in the metal resulting in field penetration into the metal. It is therefore relevant to quantify these effects in order to determine fundamental limits of the vertical field confinement of the propagating plasmon.

Before we discuss a more rigorous treatment of the  metal non-local effects, we provide a qualitative picture \cite{Luo2013}, where the non-local metal permittivity (NMP) was modelled as a dielectric shell surrounding a local metal permittivity (LMP) bulk material.
Applying this model to our system we used a perfect conductor model (i.e. with zero damping) for the the bulk metal and the thickness $s$ of a uniform dielectric spacer is used as a fitting parameter.
Simulations for this case, with an effective dielectric thickness of 3~nm (Fig~\ref{fig:meas2}C), show good agreement with the experiment (Fig~\ref{fig:meas2}B).
Even though this result conflicts with the estimated spacer thickness ($0.7$~nm), the non-local effect can be understood by the fact that the electromagnetic field does penetrate more into the metal for smaller $s$ (see Figs. 4 and S13) increasing the effective $s$.

\begin{figure}
    \includegraphics[width=0.5\textwidth]{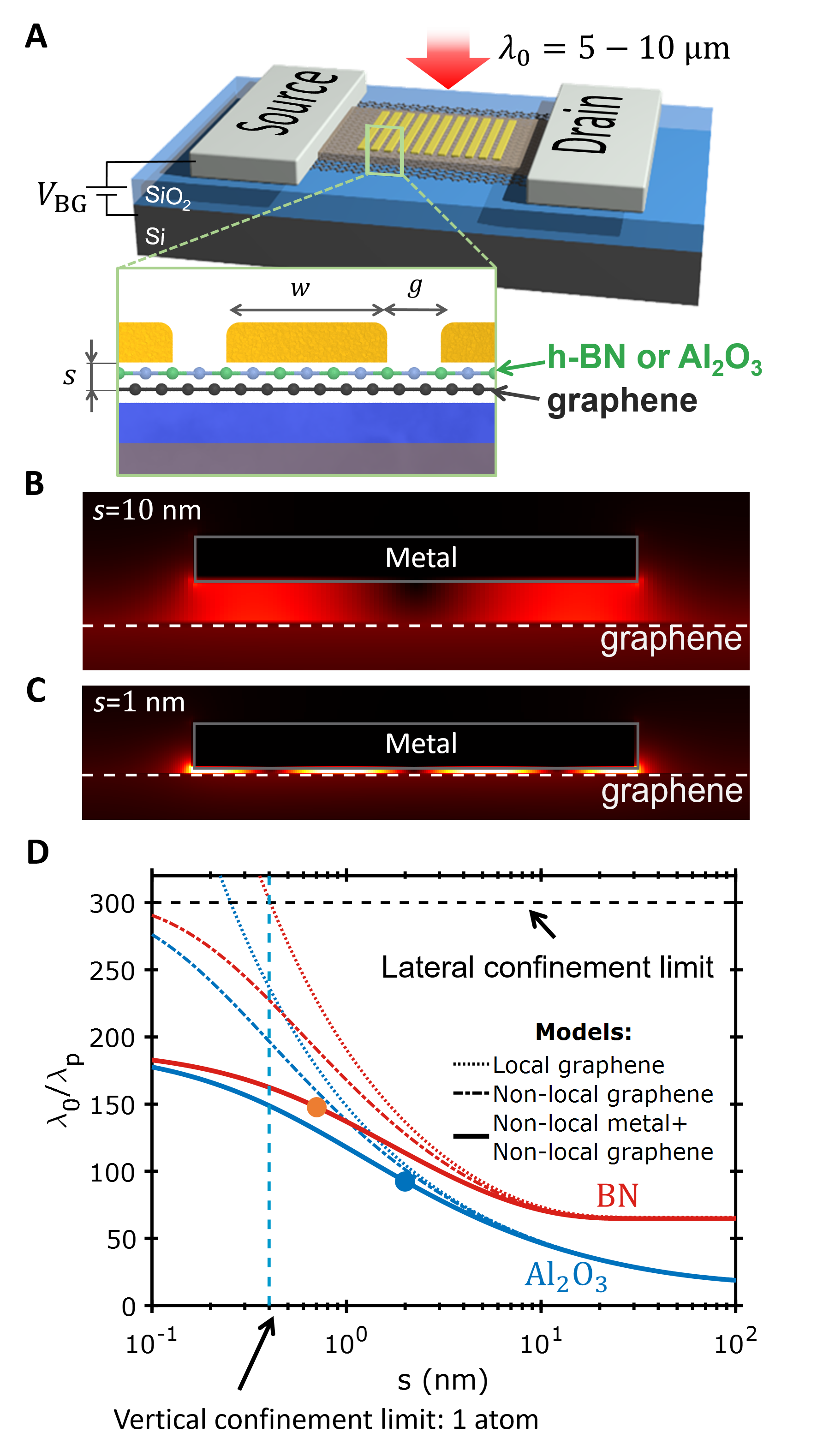}
    \caption{\textbf{Device design for probing ultimate  plasmon confinement limits.}
    (\textbf{A}) Graphene is encapsulated in a dielectric (few nm thick \ce{Al2O3} or monolayer \ce{h-BN}),  and covered by an array of gold rods. 
 A gate voltage $V_{\rm{BG}}$ is applied between  Si and graphene in order to control the Fermi energy of the graphene $E_{\rm{F}}$. Bottom left inset: Schematic cross-section of the device.  
(\textbf{B,C}) Simulated plasmonic field magnitude profiles for  metal-graphene separation $s$ of $10$ and $1$~nm.
    (\textbf{D}) Simulated plasmon wavelength $\lambda_p$ as a function of metal-graphene spacer $s$ for the two materials used in the experiments ($\lambda_0=8~\mu m$ and $E_F=0.54~eV$).       The vertical dashed line refers to the fundamental limit: a monolayer h-BN spacer. Colored circles correspond to the two sets of devices discussed in the main text.   The dotted lines represent the model where the metal was considered as a perfect conductor in combination with the local graphene conductivity model.  The dash-dotted lines represent the non-local graphene conductivity model (obtained from the random-phase-approximation), but still metal as a perfect conductor. The solid lines represent the model where non-local optical response for both metal and graphene are considered.}
    \label{fig:method1}
\end{figure}

\begin{figure}
    \includegraphics[width=0.5\textwidth]{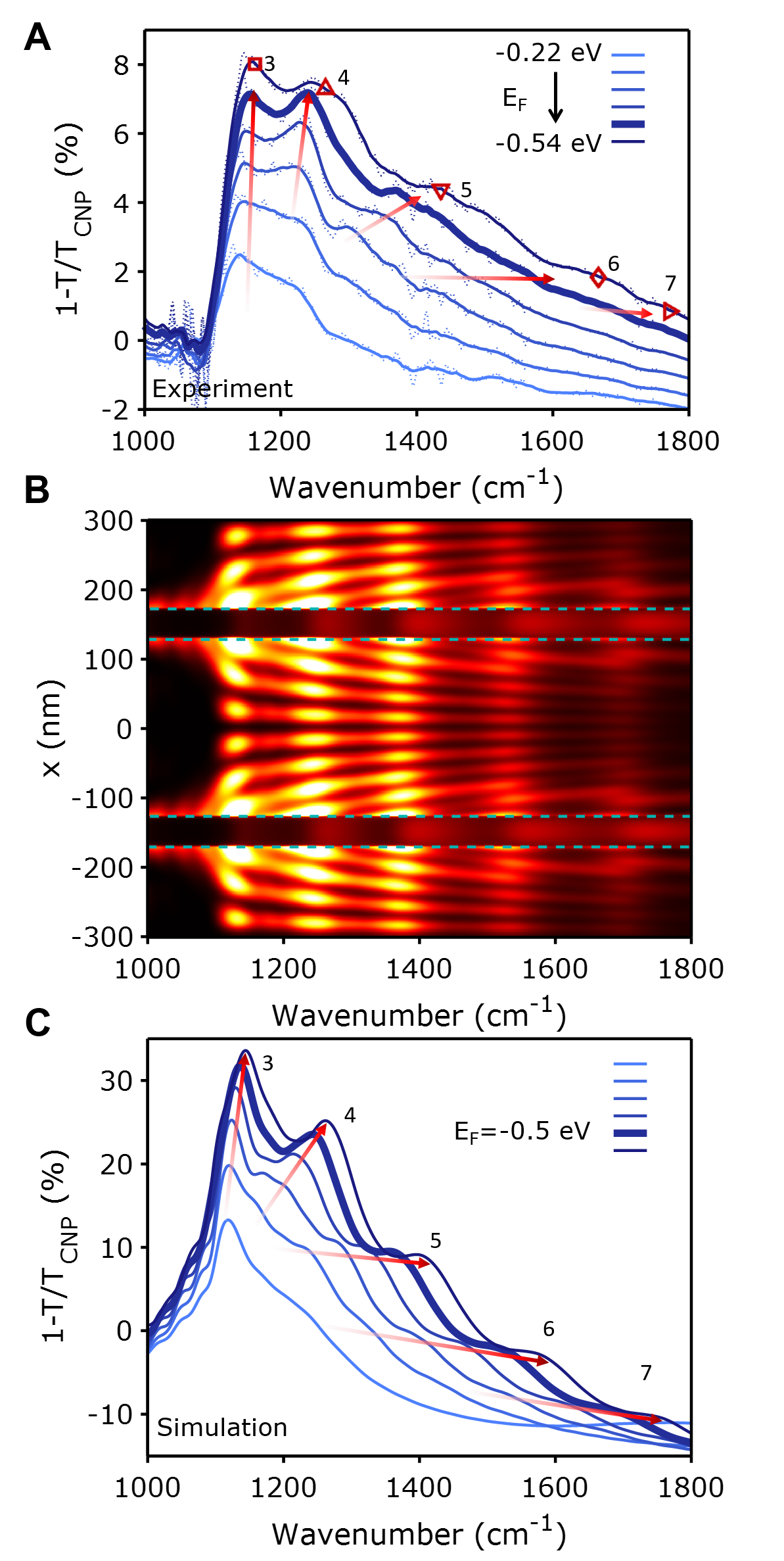}
    \caption{\textbf{Resonant excitation of 2~nm confined Fabry-P\'erot SGP modes.}
    (\textbf{A}) Gate dependent FTIR extinction spectra referenced to the charge neutrality point (CNP) for metal rod width  $w=$ 256~nm, gap $g=$ 44~nm and 2~nm \ce{Al2O3} spacer between graphene and the metal. For increasing $E_{\rm{F}}$, the resonances increase in intensity, blue-shift and high order modes become visible.    (\textbf{B})  Simulation of the electric field intensity at the graphene and along $x$, which is  along the short-axis of the rods, for the same device geometry as in panel A, with $E_{\rm{F}}$=0.5~eV. The model considers the local response of the gold and the local response of graphene. The colorscale is linear.    (\textbf{C}) Simulated extinction for the device geometry as in panel A, and same model as for panel B.}
    \label{fig:meas1}
\end{figure}

\begin{figure}
    \includegraphics[width=1\textwidth]{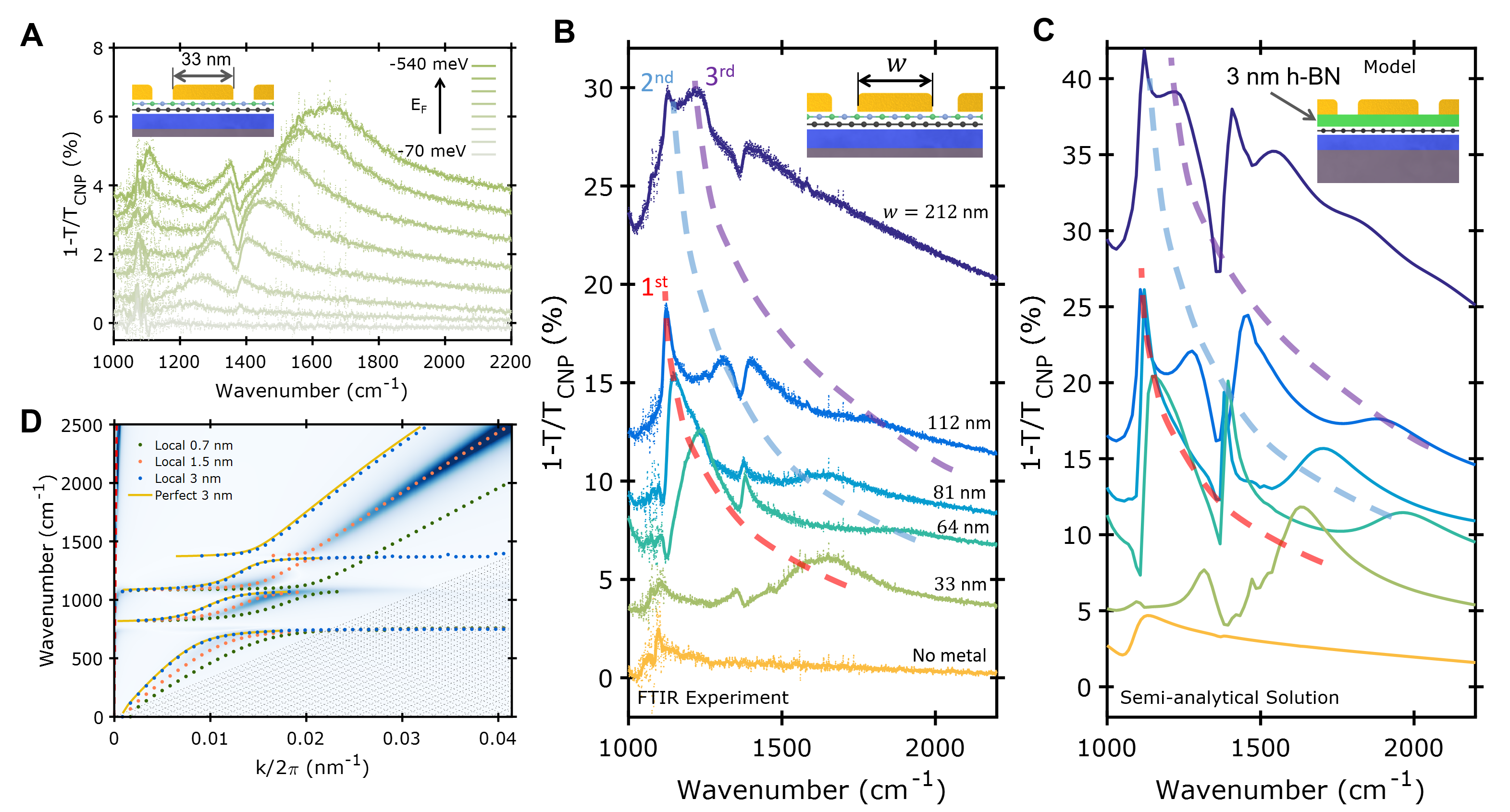}
    \caption{\textbf{Single mode and Fabry-P\'erot SGP modes confined to a monolayer \ce{h-BN}.}
    (\textbf{A}) Extinction spectra for  $E_{\rm{F}}$ ranging from 70 to 540 meV, and fixed $w\approx33$~nm, gap $g\approx37$~nm.
    (\textbf{B}) Extinction spectra for $w$ ranging from $33$~nm to $212$~nm, and fixed $g=38\pm4$~nm, $E_{\rm{F}}$=540 meV. Dashed lines are guides to the eye showing the evolution of each resonance with $w$.   Inset: monolayer device schematic for data in panels A and B.       (\textbf{C}) Simulated extinction spectra where the non-local metal effects are accounted for by modelling a perfectly conducting metal but an effective thicker $3$~nm \ce{h-BN} spacer.  Inset: Model schematic.    (\textbf{D}) Plasmon dispersion relation for a (continuous) \ce{SiO2}/Graphene/\ce{h-BN}/Metal/Air heterostructure.    Dotted curves correspond to local metal response (with non-zero loss, see \cite{sup2} section 3) and are plotted for \ce{h-BN}  thickness of $0.7$~nm (green), $1.5$~nm (orange) and $3$~nm (blue).  The solid yellow curve corresponds to \ce{h-BN}  thickness of $3$~nm (yellow) and modelling the metal as perfectly conducting.  The blue color gradient represents the loss function of the heterostructure for $0.7$~nm thick \ce{h-BN} with non-local metal (titanium) and non-local graphene response. This illustrates that accounting for non-locality comes down to adding an extra spacer thickness of $\sim2$~nm to a model that considers only the local metal response. }
    \label{fig:meas2}
\end{figure}

The excited plasmon modes associated with the extinction peaks in Figs~\ref{fig:meas1} and~\ref{fig:meas2} can be seen as a combination of SGPs under the metal and unscreened graphene plasmons in the gap region with their relative contribution depending on the geometry.
Nevertheless, the calculated field profiles in  Fig.~\ref{fig:method1}b,c and the scaling of the plasmon resonance peak energy with $w$  clearly show that the electric field (associated with the plasmon modes) is mostly confined between the metal and the graphene. This confinement is also consistent with the observation that the plasmon resonances shift most significantly when varying $w$ compared to a change of the gap $g$ (see \cite{sup2} section 3) and the fact that the system can be seen as a plasmonic crystal (analogous to photonic crystals) \cite{Lin2017}.

To further explore that a thicker dielectric in the LMP model results in non-local effects in the metal, the dispersion relation of the SGP modes is shown Fig~\ref{fig:meas2}D.
The dispersion relation for a fully NMP model is offset to higher energies compared to a LMP model accounting for the proper h-BN thickness ($0.7$~nm, green dotted line).
Accounting for a thicker dielectric spacer used to fit our data ($s=1.5$~nm), the dispersion curves shift down and overlaps with the NMP model.  Qualitatively, this is in agreement with our findings. An additional shift is expected for a periodic structure instead of a continuous one as calculated in Fig~\ref{fig:meas2}D, because of a small coupling between the modes below the rods.

\begin{figure}
    \includegraphics[width=1\textwidth]{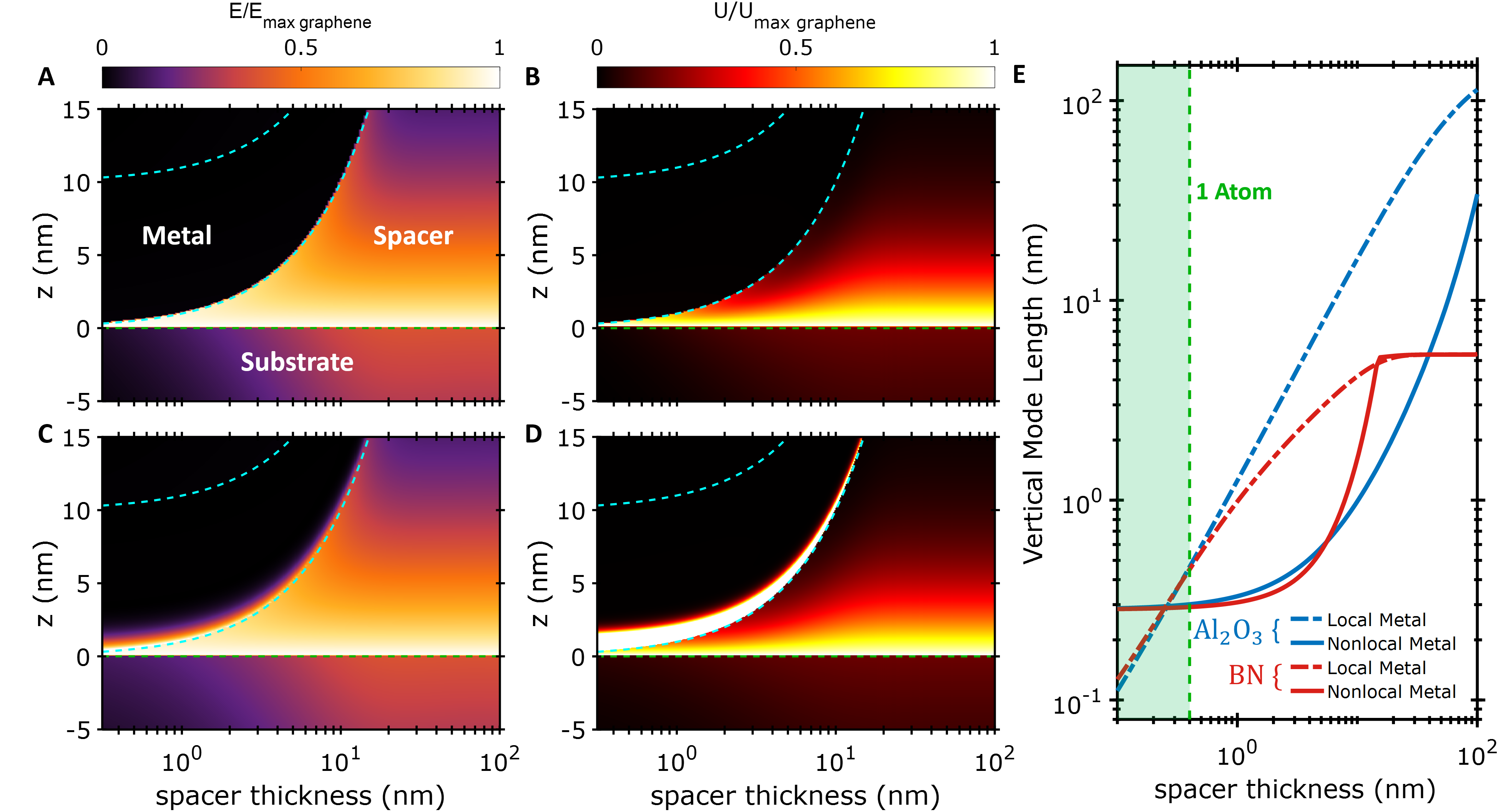}
    \caption{\textbf{Energy density and field confinement.}
    (\textbf{A}) Electric field magnitude distribution of the plasmons associated to a continuous heterostructure of air/ Ti/\ce{h-BN}/graphene/\ce{SiO_2}  as function of h-BN  thickness for LMP model (A) and NMP model (C).
    The top and bottom metal limits are depicted by cyan dashed lines and graphene is located at $z=0$.
    Normalization by the maximum electric field strictly above graphene shows the confinement and screening effects.
    (\textbf{B}) same as (C) for the energy density.
    (\textbf{E}) Vertical field confinement for both types of dielectrics as a function of the spacer thickness for local (dotted) and non-local (solid) metal permittivity.}
    \label{fig:modevolume}
\end{figure}

We can now evaluate the ultimate limit on the plasmonic vertical field and mode volume confinement. We have calculated the electric field intensity and energy density distribution for different spacer thicknesses (Fig.~\ref{fig:modevolume}) and considered both LMP and NMP models.
We used the definitions in section 5 of \cite{sup2}, taking into account dispersion effects.
Inspecting first the electric field magnitude obtained from the LMP model and normalized by the maximum of the graphene plasmon (Fig.~\ref{fig:modevolume}A,B), we find that most of it is confined between the graphene and the metal and the penetration of the field inside the the metal is negligible. 
On the other hand, considering metal non-local effects (NMP model) offers a more complete picture of the physics of the electrons that accumulate at the metal surface in order to screen the electromagnetic field \cite{teperik2013robust}.
When the out of plane wavector is increased for thinner spacers, the electrons start suffering from Pauli and electrostatic repulsion.
This effect results into a saturation of the electron density and leads to field penetration into the metal, as the metal screening capability is reduced \cite{David2014}. The penetration of the field into the metal becomes significant for $s$ below 3 nm, although the field remains maximum in the spacer region.

This non-local field penetration limits the vertical mode length, which is defined by the ratio of the energy density integrated over the out-of-plane coordinate $z$ to the maximum of the field intensity in the region of the spacer \cite{ruppin2002electromagnetic,maier2006plasmonic}:
\begin{equation}
L=\frac{\int u_E(z) dz}{\max{u_E(z)}}.
\label{eq:vertical_length}
\end{equation}
The energy density distribution (Fig. 4D) peaks at the metal surface for a NMP model as a consequence of charge accumulation and the combination of high field and permittivity values, which is in contrast to the LMP (Fig.~\ref{fig:modevolume}C).
Below a certain spacer thickness, the energy density at the metal surface becomes larger than the energy density in the dielectric region near graphene.
At this point the transition of the maximum energy density from graphene to the metal dominates the vertical mode length (Fig.~\ref{fig:modevolume}E). The out-of-plane confinement   is calculated from Eq.~\eqref{eq:vertical_length} and the energy density distribution that is shown in Fig.~\ref{fig:modevolume}E. While for LMP, the vertical mode length can be made arbitrarily small, the non-local metal behaviour limits the vertical mode length to about 0.3 nm.  This corresponds to a plasmonic mode confinement down to the atomic scale, as confirmed by our experiments. 

These results raise the question why the plasmonic mode for $s=0.7$ nm, which  penetrates the metal (due to non-local response), does not lead to strong (over)damping by Landau damping? For metals, the strongest confinement normal to the surface is limited by direct excitation of electron-hole pairs (Landau damping), accounted for by the non-local response function \cite{bozhevolnyi2016quantum}. A quantitive analysis of the metallic non-local corrections to the dynamic response and damping \cite{teperik2013robust,Christensen2017a} trough the Feibelman parameters revealed that these effects are much stronger for  $\omega$ approaching  $\omega_p$. Interestingly, for $\omega \ll \omega_p$, the imaginary part of the Feibelman parameters approaches zero, as the phase space for electon-hole excitations decreases with $\omega$. The real part of the Feibleman parameters do not approach zero for $\omega \ll \omega_p$.  Thus, for our experimental conditions, the metal non-local effects are relevant as the electrons in the metal cannot perfectly screen the field, but the additional damping from electronic excitations in the metal is weak.

In terms of  ultimate mode volume limit, we remark that a reduction of vertical confinement reduces also the lateral propagating plasmon wavelength  (Fig.~\ref{fig:method1}D).  
While our experiment discloses the fundamental limits of the vertical mode length of $\lambda_0/26000$, it also allows us to estimate the complete mode volume confinement $V_{p}$ with respect to the volume associated with the extent of the free photons $V_0=\lambda_0^3$. If instead of an array (as in this work) a single resonant structure (e.g. a metal disc on graphene, with a monolayer h-BN spacer) was built using our approach, a vertical mode length down to $0.3$~nm and plasmon wavelenth $\lambda_p=\lambda_0/170$ (see Fig.~\ref{fig:method1}D) would be attainable.
Using these values (for $\lambda_0=8~\mu m$) that include the full non-local response for both graphene and metal, this would correspond to a mode volume of $V_p=664~nm^3$ and mode volume ratio $V_0/V_p\approx10^{-9}$, which is at least two orders of magnitude smaller than reported so far with graphene nanoribbons \cite{Brar2013}, and sufficiently high to explore new regimes of light-matter interactions such as the ultra-strong coupling regime. 

Our results show that 2D-material heterostructures can be considered as a powerful toolbox for nano-photonics with vertical sub-nanometer precision. We use this to combine efficient coupling between radiation and plasmons with extreme confinement,
 and found that propagating graphene plasmons can be confined down to 0.3 nm and drive a low-loss non-local metal response.  
The metallic structure can also be used as a nearby efficient gate\cite{Popov2011} which provides a route to low-voltage applications to study molecular sensing with even higher resolution, to enhance non-linear effects, or to design photodetectors with plasmon enhanced sensitivity and tunability.




\begin{thebibliography}{10}

\bibitem{geim2013}
A.~K. Geim, I.~V. Grigorieva, {\it Nature\/} {\bf 499}, 419 (2013).

\bibitem{Novoselov2016}
K.~Novoselov, A.~Mishchenko, A.~Carvalho, A.~C. Neto, {\it Science\/} {\bf 353}
  (2016).

\bibitem{low2017polaritons}
T.~Low, {\it et~al.\/}, {\it Nature materials\/} {\bf 16}, 182 (2017).

\bibitem{basov2016polaritons}
D.~Basov, M.~Fogler, F.~G. de~Abajo, {\it Science\/} {\bf 354} (2016).

\bibitem{Woessner2014}
A.~Woessner, {\it et~al.\/}, {\it Nature Mater.\/} {\bf 14}, 421 (2015).

\bibitem{principi2011}
A.~Principi, R.~Asgari, M.~Polini, {\it Solid State Communications\/} {\bf
  151}, 1627 (2011).

\bibitem{Popov2011}
V.~V. Popov, {\it Journal of Infrared, Millimeter, and Terahertz Waves\/} {\bf
  32}, 1178 (2011).

\bibitem{Alonso-Gonzalez2017}
P.~Alonso-Gonz{\'a}lez, {\it et~al.\/}, {\it Nature nanotechnology\/} {\bf 12},
  31 (2017).

\bibitem{Lundeberg2017}
M.~B. Lundeberg, {\it et~al.\/}, {\it Science\/} {\bf 357}, 187 (2017).

\bibitem{Stockman2004}
M.~I. Stockman, {\it Physical Review Letters\/} {\bf 93}, 137404 (2004).

\bibitem{Bozhevolnyi2005}
S.~I. Bozhevolnyi, V.~S. Volkov, E.~Devaux, T.~W. Ebbesen, {\it Physical Review
  Letters\/} {\bf 95}, 046802 (2005).

\bibitem{Oulton2008}
R.~F. Oulton, V.~J. Sorger, D.~A. Genov, D.~F.~P. Pile, X.~Zhang, {\it Nature
  Photonics\/} {\bf 2}, 496 (2008).

\bibitem{Dionne2006}
J.~A. Dionne, L.~A. Sweatlock, H.~A. Atwater, A.~Polman, {\it Physical Review
  B\/} {\bf 73}, 035407 (2006).

\bibitem{miyazaki2006}
H.~T. Miyazaki, Y.~Kurokawa, {\it Physical review letters\/} {\bf 96}, 097401
  (2006).

\bibitem{bozhevolnyi2016quantum}
S.~Bozhevolnyi, L.~Martin-Moreno, F.~Garcia-Vidal, {\it Quantum Plasmonics\/},
  Springer Series in Solid-State Sciences (Springer International Publishing,
  2016).

\bibitem{Ciraci2012}
C.~Ciraci, {\it et~al.\/}, {\it Science\/} {\bf 337}, 1072 (2012).

\bibitem{Benz2016}
F.~Benz, {\it et~al.\/}, {\it Science\/} {\bf 354}, 726 (2016).

\bibitem{teperik2013robust}
T.~V. Teperik, P.~Nordlander, J.~Aizpurua, A.~G. Borisov, {\it Physical review
  letters\/} {\bf 110}, 263901 (2013).

\bibitem{Brar2013}
V.~W. Brar, M.~S. Jang, M.~Sherrott, J.~J. Lopez, H.~A. Atwater, {\it Nano
  Letters\/} {\bf 13}, 2541 (2013).

\bibitem{Yan2013c}
H.~Yan, {\it et~al.\/}, {\it Nature Photonics\/} {\bf 7}, 394 (2013).

\bibitem{sup2}
See the supplementary materials.

\bibitem{Lin2017}
I.-T. Lin, C.~Fan, J.-M. Liu, {\it IEEE Journal of Selected Topics in Quantum
  Electronics\/} {\bf 23}, 144 (2017).

\bibitem{Fang2013}
Z.~Fang, {\it et~al.\/}, {\it ACS Nano\/} {\bf 7}, 2388 (2013).

\bibitem{Kim2012}
K.~K. Kim, {\it et~al.\/}, {\it Nano Letters\/} {\bf 12}, 161 (2012).

\bibitem{Park2014}
J.~H. Park, {\it et~al.\/}, {\it ACS Nano\/} {\bf 8}, 8520 (2014).

\bibitem{Brar2014}
V.~W. Brar, {\it et~al.\/}, {\it Nano Lett.\/} {\bf 14}, 3876 (2014).

\bibitem{Luo2013}
Y.~Luo, A.~I. Fernandez-Dominguez, A.~Wiener, S.~A. Maier, J.~B. Pendry, {\it
  Physical Review Letters\/} {\bf 111}, 1 (2013).

\bibitem{David2014}
C.~David, F.~J.~G. de~Abajo, {\it ACS-Nano\/} {\bf 8}, 9558 (2014).

\bibitem{ruppin2002electromagnetic}
R.~Ruppin, {\it Physics letters A\/} {\bf 299}, 309 (2002).

\bibitem{maier2006plasmonic}
S.~A. Maier, {\it Optics Express\/} {\bf 14}, 1957 (2006).

\end{thebibliography}

\section*{Acknowledgments}
The authors thank Gerasimos Konstantatos and Valerio Pruneri for the intensive use of their respective FTIRs, very insightful discussions with Marco Polini, Thomas Christensen, Asger Mortenson and Javier Aizpurua on non-local effects and with Achim Woessner on simulation and modelling of graphene acoustic plasmons modes.
{\bf Funding:} We acknowledge financial support from the Spanish Ministry of Economy and Competitiveness, through the Severo Ochoa Programme for Centres of Excellence in R\&D (SEV-2015-0522), support by Fundacio Cellex Barcelona, the Mineco grants Ramon y Cajal (RYC-2012-12281), Plan Nacional (FIS2013-47161-P and FIS2014-59639-JIN), and the Government of Catalonia trough the SGR grant (2014-SGR-1535). Furthermore, the research leading to these results has received funding from the European Union H2020 Programme under grant agreement no. 604391 Graphene Flagship, the ERC starting grant (307806, CarbonLight) and project GRASP (FP7-ICT-2013-613024-GRASP). D.A.I. acknowledges the FPI grant BES-2014-068504. N.M.R.P. and E.J.C.D acknowledge support from the Portuguese Foundation for Science and Technology (FCT) in the framework of the Strategic Financing UID/FIS/04650/2013. This work was supported in part by the Center for Excitonics, an Energy Frontier Research Center funded by the U.S. Department of Energy, Office of Science, Office of Basic Energy Sciences under Award No. DE-SC0001088, and the Army Research Office (grant number 16112776).  J.-Y.H. and J.K. acknowledge support from the US AFOSR FATE MURI, Grant No. FA9550-15-1-0514.
{\bf Author contributions}: F.H.L.K, D.A.I. and S.N. conceived the idea; E.J.C.D and N.M.R.P. developed the analytical model; S.N., C.P., J.O., D.E. and D.A.I. fabricated the devices; S.N., R.P. and D.A.I. performed measurements; D.A.I., M.L., I.E., and S.N. performed data analysis; J.Y.H. and J.K. provided h-BN; D.A.I., S.N., E.J.C.D., N.M.R.P, I.E., D.E. and F.H.L.K wrote the manuscript; D.E. and F.K. supervised the project.
{\bf Competing interests}: None of the authors have competing interests. 
{\bf Data and materials availability}: All data needed to evaluate the conclusions in the paper are present in the paper and/or the Supplementary Materials

\section*{Supplementary materials}
Materials and Methods\\
Supplementary Text\\
Figs. S1 to S17\\
References \textit{(32-46)}

\clearpage


\end{document}